\documentclass{llncs}

\usepackage[nolineno, noindent]{lgrind}
\usepackage{url}
\usepackage{verbatim}

\usepackage[dvips]{graphics}
\usepackage{epsfig}
\usepackage{color}
\usepackage{graphicx}

\newcommand{\ang}[1]{\em{#1}}

\def\LGsize{\scriptsize}

\makeatletter
\def\ifundefined{\@ifundefined}
\makeatother

\newcommand{\CAMIL}{OCamIL}
\newcommand{\DOTNET}{.NET}
\newcommand{\CSHARP}{C\#}

\newcommand{\FSHARP}{F\#}

\newcommand{\JAVA}{Java}

\newcommand{\OComplete}{Objective Caml}
\newcommand{\OC}{O'Caml}

\newcommand{\IDL}{IDL}

\newcommand{\OJACARE}{{\tt{O'Jacar\'e}}}
\newcommand{\OJACARED}{{\tt{O'Jacar\'e.net}}}

\newcommand{\forget}[1]{}

\begin{document}

\frontmatter
\pagestyle{headings}
\addtocmark{{\CAMIL} Toplevel}

\mainmatter

\title{Mixing the {\OComplete} and {\CSHARP} Programming Models in the 
{\DOTNET} Framework}
\titlerunning{Mixing {\OC} and {\CSHARP}}

\author{Emmanuel Chailloux\inst{1} \and Gr\'egoire Henry\inst{1} \and Rapha\"{e}l Montelatici\inst{2}}



\institute{
Equipe Preuves, Programmes et Syst{\`e}mes (CNRS UMR 7126) \\
         Universit\'e Pierre et Marie Curie (Paris 6) - 
         4 place Jussieu, 75005 Paris, France\\
\email{Emmanuel.Chailloux@pps.jussieu.fr,Gregoire.Henry@pps.jussieu.fr}\\
\and
Equipe Preuves, Programmes et Syst{\`e}mes (CNRS UMR 7126) \\
 Universit\'e Denis Diderot (Paris 7) -  2 place Jussieu, 75005 Paris, France\\
\email{Raphael.Montelatici@pps.jussieu.fr}
}

\maketitle



\begin{abstract}
We present a new code generator, called {\OJACARED}, to inter-operate between {\CSHARP}~and {\OComplete} through their object models. 
{\OJACARED} defines a basic {\IDL}~{\ang (Interface Definition Language)} that describes classes and interfaces in order
to communicate between {\OComplete} and {\CSHARP}. 
{\OJACARED} generates  all needed wrapper classes  and takes advantage of static type checking in both worlds. 
Although the {\IDL} intersects these two object models, {\OJACARED} allows to combine features from both.

\end{abstract}



\section{Introduction}

The {\DOTNET}
platform claims to be a melting  pot that allows the integration of 
different languages in a common framework, sharing a common type system, CTS, 
and a runtime environment, CLR ({\emph{Common Language Runtime}}). 
Each {\DOTNET} compiler generates portable MSIL byte-code 
({\emph{MicroSoft Intermediate Language}}). 
By assuming compliance to the CTS type system, components inter-operate safely.

The {\DOTNET} framework is actually well suited for object-oriented languages which have an object model close to the one of {\CSHARP} or {\JAVA}. 
Unfortunately, languages with other kinds of object models, type systems 
or supporting different programming paradigms (such as functional programming \ldots) 
do not fit in {\DOTNET} as well as {\CSHARP} does. 
Writing {\DOTNET} compilers for them requires much more efforts.

However, the {\DOTNET} framework still gives us a good opportunity to experiment inter-operability between two languages as different as {\OComplete}\cite{ocaml-refman:2002} 
(shortened as {\OC}) and {\CSHARP}.
{\OC} is an  ML dialect: it is a functional/imperative statically typed language, 
featuring parametric polymorphism, an exception mechanism, an object layer and parameterized modules. 
By achieving inter-operability, each language gains access to a wider set of libraries and programmers
take advantage of a richer programming model.

We use the experimental {\CAMIL} compiler\cite{mcp-ocamil:2004}, which compiles the whole {\OC} distribution (including toplevel) to {\DOTNET} managed code.
We intend to communicate between {\OC} and {\CSHARP} by means of their respective object models. 
Difficulties arise because neither the type system nor the object model of {\OC} natively fit in the CTS.
{\OC} objects cannot be directly compiled to CTS objects. 
Communication cannot be direct: {\CSHARP} and {\OC} objects have to be interfaced.
We use an {\IDL} (Interface Description Language) and a code generator called {\OJACARED}.
It is based on our previous work\cite{chailloux-henry:2004} on inter-operability of {\OC} and {\JAVA}.

We first describe the {\OC} object model and compare it to the {\CSHARP} model.
We then introduce {\OJACARED}, using a small example as an illustration of its features.
The last section is dedicated to expressiveness issues.
We show that the combination of two object models allows to take advantage of features of both, 
and also discuss the current limitations of {\OJACARED}, giving hints on how they can be solved in further developments.

\section{Comparing Object Models}
 
{\OC} is a statically typed language based on a functional and imperative kernel.
It also integrates a class-based object-oriented extension in its type system,
for which inheritance relation and subtyping relation for classes are well distinguished\cite{remy98objective}. 
One key feature of {\OC} type system is type inference.
The programmer does not annotate programs with typing indications:
the compiler gives each expression the most general type it cans.\\

\noindent A class declaration defines: 
\begin{itemize} 
\item a new type abbreviation of an object type,
\item a constructor function to build class instances. 
\end{itemize}
An object type is 
characterized by the name and the type of its methods. 
For instance, the following type can be inferred for class instances which declare {\tt{moveto}} and {\tt{toString}} methods: 
\begin{center}
\begin{minipage}{12cm}
\begin{center}
\verb+< moveto : (int * int) -> unit; toString : unit -> string >+
\end{center}
\end{minipage}
\end{center}
At each method call site, static typing checks that the type of the receiving instance is
an object type and that it contains the relevant method name with a compatible type. 
The following example is correct if the class {\tt{point}} defines (or inherits) 
a method {\tt{moveto}} expecting a couple of integers as argument. 
Within the {\OC} type inference,
the most general types given to objects are expressed by means of ``open'' types ({\tt{<..>}}). 
The function {\tt{f}} can be used with any object having a method {\tt{moveto}} (\verb+'a+ denotes a universally quantified type variable): 

\begin{center}
\begin{tabular}{|c|c|}
\hline
method call & functional-object style \\ \hline
\begin{minipage}{3.5cm}
\vspace{2pt}
\LGinlinefalse \lgrinde
\index{}\Proc{}\L{\LB{\K{let}_\V{p}_=_\K{new}_\V{point}(\N{1},\N{1});;}}
\L{\LB{\V{p}\V{\#moveto}(\N{10},\N{2});;}}
\endlgrinde\LGend
\end{minipage}
& 
\begin{minipage}{7cm}
\vspace{2pt}
\LGinlinefalse \lgrinde
\index{}\Proc{}\L{\LB{\V{\#}_\K{let}_\V{f}_\V{o}_=_\V{o}_\V{\#}_\V{moveto}_(\N{10},\N{20});;}}
\L{\LB{\K{val}_\V{f}_:_\<_\V{moveto}_:_\V{int}_*_\V{int}_\-\!\>_\V{{'}a};_.\,._\>_\-\!\>_\V{{'}a}}}
\endlgrinde\LGend
\end{minipage} \\ \hline
\end{tabular}
\end{center}
\newpage
\noindent Here are some of {\OC} object model most important characteristics:
\begin{itemize}
\item Class declarations allow multiple inheritance and parametric classes. 
\item Method overloading is not supported. 
\item The methods binding is always delayed.
\end{itemize}

The {\CSHARP} language model is well known and will not be described here. 
We compare its main features with {\OC} 
in the following table:
%

{\small{
\begin{center}
\begin{tabular}{|l|c|c||l|c|c|}\hline
Features & {\CSHARP} & {\OC}  & Features & {\CSHARP} & {\OC} \\ \hline 
classes & $\surd$ & $\surd$ &  inheritance $\equiv$ sub-typing? & yes & no  \\ \hline
late binding &  $\surd$ & $\surd$  & overloading & $\surd$ & 3 \\ \hline
early binding & $\surd$ & 1  &  multiple inheritance & 4 & $\surd$\\ \hline
static typing & $\surd$ & $\surd$  &  parametric classes& 5 & $\surd$\\ \hline
dynamic typing & $\surd$ & 2 &  packages/modules & 6 & 6\\ \hline
sub-typing & $\surd$ & $\surd$  &  & & \\ \hline
\end{tabular}
\end{center}
}}

\label{fig-modeles}


\indent1) static methods are global functions of a {\OC} module and class variables are global declarations;\\
\indent2) no downcast in {\OC} 
(only available in the coca-ml~\cite{chailloux:2002b} extension);\\
\indent3) no overloading in {\OC} but the type of {\tt{self}} can appear in the type of a method eventually overridden in a subclass;\\
\indent4) no multiple inheritance for {\CSHARP} classes, only for interfaces;\\
\indent5) generics\cite{KenSyme01} are expected in {\CSHARP} 2.0;\\
\indent6) simple modules of {\OC} correspond to public parts of {\CSHARP} namespaces; there is no parameterized modules in {\CSHARP}.\\

The intersection of these two models corresponds to a basic class-based
 language, where method calls are delayed, and inheritance and subtyping relations are equivalent. 
Concerning type system, there is no overloading and no binary methods.
For the sake of simplicity, there is no multiple inheritance nor parametric classes. 
This model inspires a basic {\IDL} for interfacing {\CSHARP} and {\OC} classes.

\section{Introducing {\OJACARED}}
\label{introducing-section}
{\OJACARED} is based on our previous work {\OJACARE} on {\OC} and {\JAVA}.
Its purpose was to use {\JAVA} objects in {\OC}. We encountered difficulties with the management of two different runtimes
({\JAVA} runtime and {\OC} runtime), especially for handling threads and garbage collection.
Adapting this work to {\CSHARP} and the {\CAMIL} implementation of {\OC} on {\DOTNET} makes things easier, 
mainly because there is only one runtime.

{\OJACARED} allows to use {\CSHARP} objects in {\OC}, and  {\OC} objects in {\CSHARP} as well.


\subsection{{\CSHARP} in {\OC}}
Our current communication model between {\OC} and {\CSHARP} affords two levels
of communication:
as the first level provides a basic encapsulation mechanism
of {\CSHARP} objects inside {\OC} objects,
the second level adds a callback mechanism that allows to override {\CSHARP} 
methods in {\OC} using late binding.

\subsubsection{Basic encapsulation}

Starting from the description of classes and interfaces in an {\IDL} file, {\OJACARED} generates 
wrappers in the target language (here, {\OC}), allowing to allocate objects
and call methods upon classes of the foreign language (here, {\CSHARP}) as if
those classes were native.

Let us illustrate this mechanism on a small example:
we want to handle two {\CSHARP} classes, {\tt{Point}} and {\tt{Colored\-Point}}.
They are described in the {\IDL} file below.
\begin{center}
\begin{tabular}{|ll|ll|} \hline
\multicolumn{4}{|c|}{File {\tt p.idl}} \\ \hline
&
\begin{minipage}{4.8cm}
\vspace{2pt}
\LGinlinefalse \lgrinde
\L{\LB{\K{package}_[\K{assembly}_\V{point}]_\V{mypack};}}
\L{\LB{}}
\L{\LB{\K{class}_\V{Point}_\{_}}
\L{\LB{}\Tab{2}{\K{int}_\V{x};_\K{int}_\V{y};_}}
\L{\LB{}\Tab{2}{}}
\L{\LB{}\Tab{2}{[\K{name}_\V{default\_point}]_\K{\<init\>}_();}}
\L{\LB{}\Tab{2}{[\K{name}_\V{point}]_\K{\<init\>}_(\K{int},\K{int});}}
\L{\LB{}\Tab{2}{\K{void}_\V{moveTo}(\K{int},\K{int});}}
\L{\LB{}\Tab{2}{\K{string}_\V{toString}();}}
\L{\LB{}\Tab{2}{\K{void}_\V{display}();}}
\L{\LB{}\Tab{2}{\K{boolean}_\V{equals}(\V{Point});}}
\L{\LB{\}}}
\endlgrinde\LGend
\vspace{2pt}
\end{minipage}
&&
\begin{minipage}{6.4cm}
\vspace{2pt}
\LGinlinefalse \lgrinde
\L{\LB{\K{interface}_\V{Colored}_\{}}
\L{\LB{}\Tab{2}{\K{string}_\V{getColor}();}}
\L{\LB{}\Tab{2}{\K{void}_\V{setColor}(\K{string});}}
\L{\LB{\}}}
\L{\LB{}}
\L{\LB{\K{class}_\V{ColoredPoint}_\K{extends}_\V{Point}_}}
\L{\LB{}\Tab{24}{\K{implements}_\V{Colored}_\{}}
\L{\LB{}\Tab{2}{[\K{name}_\V{default\_colored\_point}]_\K{\<init\>}_();}}
\L{\LB{}\Tab{2}{[\K{name}_\V{colored\_point}]_\K{\<init\>}_(\K{int},\K{int},\K{string});}}
\L{\LB{}\Tab{2}{}}
\L{\LB{}\Tab{2}{[\K{name}_\V{equals\_pc}]_\K{boolean}_\V{equals}(\V{ColoredPoint})}}
\L{\LB{\}}}
\endlgrinde\LGend
\vspace{2pt}
\end{minipage} \\ \hline
\end{tabular}
\end{center}

The {\IDL} syntax borrows from {\JAVA} syntax and is extended with attributes 
(i.e. for name-aliasing because overloading is not allowed).

For the {\tt p.idl} file, {\OJACARED} generates an {\OC} module, named  
{\tt p.ml}, that contains:
\begin{itemize}
\item 3 class types: {\tt csPoint}, {\tt csColored}
 and {\tt csColoredPoint} ;
\item 3 wrapper classes exposed with previous class types ;
\item 4 constructors that allocate and initialize {\CSHARP} objects, wrapping them inside the previous classes.
\end{itemize}

\noindent An example of use is illustrated in an {\OC} toplevel session below
(the \verb+equals+ method compares two objects using their instance variables \verb+x+ and \verb+y+).
\begin{center}
\begin{tabular}{|ll|ll|} \hline
\multicolumn{4}{|c|}{{\OC} toplevel session} \\ \hline
&
\begin{minipage}{5.8cm}
{\scriptsize{
\LGinlinefalse \lgrinde
\L{\LB{\K{\#}_\K{open}_\V{P}\K{;;}}}
\L{\LB{\K{\#}_\K{let}_\V{p}_=_\K{new}_\V{point}_\N{1}_\N{2}\K{;;}}}
\L{\LB{\N{val}_\N{p}_:_\N{point}_=_\N{\<obj\>}}}
\L{\LB{\K{\#}_\K{let}_\V{p2}_=_\K{new}_\V{default\_point}_\V{()}\K{;;}}}
\L{\LB{\N{val_p2_:_default\_point_=_\<obj\>}}}
\L{\LB{\#_\K{let}_\V{pc}_=_\K{new}_\V{colored\_point}_\N{3}_\N{4}_\N{"blue"}\K{;;}}}
\L{\LB{\N{val_pc_:_colored\_point_=_\<obj\>}}}
\L{\LB{\K{\#}_\K{let}_\V{pc2}_=_\K{new}_\V{default\_colored\_point}_\V{()}\K{;;}}}
\L{\LB{\N{val_pc2_:_default\_colored\_point_=_\<obj\>}}}
\L{\LB{\K{\#}_\V{p}\K{\#}\V{toString}_\V{()}\K{;;}}}
\L{\LB{\N{-_:_string_=_"(1,2)"}}}
\endlgrinde\LGend
}}
\end{minipage}
&&
\begin{minipage}{4cm}
{\scriptsize{
\LGinlinefalse \lgrinde
\L{\LB{\K{\#}_\V{pc}\K{\#}\V{toString}_\V{()}\K{;;}}}
\L{\LB{\N{-_:_string_=_"(3,4):blue"}}}
\L{\LB{\K{\#}_\V{p}\K{\#}\V{equals}_\V{(}\V{pc}_\K{:\>}_\V{csPoint}\V{)}\K{;;}}}
\L{\LB{\N{-_:_bool_=_false}}}
\L{\LB{\K{\#}_\V{pc}\K{\#}\V{moveTo}_\N{1}_\N{2}\K{;;}}}
\L{\LB{\N{-_:_unit_=_()}}}
\L{\LB{\K{\#}_\V{pc}\K{\#}\V{equals}_\V{p}\K{;;}}}
\L{\LB{\N{-_:_bool_=_true}}}
\L{\LB{\K{\#}_\V{pc}\K{\#}\V{equals\_pc}_\V{pc2}\K{;;}}}
\L{\LB{\N{-_:_bool_=_false}}}
\endlgrinde\LGend
}}
\end{minipage}\\ \hline
\end{tabular}
\end{center}

The type coercion operator {\verb+:>+} allows to consider the type of an object as a supertype, according to the subtyping relation.

\subsubsection{Callback mechanism}
\label{callback-section}
We go on with the previous example.
The {\CSHARP} implementation of the {\tt toString} method of class 
{\tt{Colored\-Point}} concatenates the results of a call to the superclass 
{\tt toString} method and a call to the {\tt{getColor}} method on itself. 
We want to redefine the {\tt getColor} method in {\OC}, and so specialize
the {\tt toString} method through late binding.

With basic encapsulation, a {\CSHARP} instance of {\tt ColoredPoint} has 
no knowledge of the {\OC} instance. 
We need a second level of communication, introduced by the 
{\tt callback} attribute :
\begin{center}
\begin{minipage}{9cm}
{\LGinlinefalse \lgrinde
\L{\LB{[\K{callback}]_\K{class}_\V{ColoredPoint}_\K{extends}_\V{Point}_\K{implements}_\V{Colored}_\{_\ldots_\}}}
\endlgrinde\LGend}
\end{minipage}
\end{center}

With this attribute, the compilation of the file {\tt{p.idl}} generates a new 
file called {\tt{ColoredPointStub.cs}} and add stub classes to the 
generated {\OC} file. 
As shown in right column of the below example, inheriting the stub in {\OC}
allows the expected behavior, whereas inheriting the wrapper 
(left column) does not!

\begin{center}
\begin{tabular}{|ll|ll|} \hline
&
\begin{minipage}{5.8cm}
{\scriptsize{
\LGinlinefalse \lgrinde
\L{\LB{\K{\#}_\K{class}_\V{wrong\_ml\_colored\_point}_\V{x}_\V{y}_\V{c}_=}} 
\L{\LB{__\K{object}}}
\L{\LB{_____\K{inherit}}}
\L{\LB{__________\V{colored\_point}_\V{x}_\V{y}_\V{c}_\K{as}_\V{super}}}
\L{\LB{_____\K{method}_\V{getColor}_\V{()}_\K{=}}}
\L{\LB{__________\N{"ML"}_\^_\V{super}\K{\#}\V{getColor}_\V{()}}}
\L{\LB{__\K{end}\K{;;}}}
\L{\LB{\N{class_wrong\_ml\_colored\_point_:}}}
\L{\LB{_____________int_-\>_int_-\>_string_-\>_csColoredPoint}}
\L{\LB{\K{\#}} \K{let} \V{wml\_cp}_=}
\L{\LB{____\K{new}_\V{wrong\_ml\_colored\_point}_\N{6}_\N{7}_\V{"green"}\K{;;}
}}
\L{\LB{\N{val_wml\_cp_:_wrong\_ml\_colored\_point_=_\<obj\>}}}
\L{\LB{\K{\#}_\V{wml\_cp}\K{\#}\V{toString}_\V{()}\K{;;}}}
\L{\LB{\N{- : string = "(6,7):green"}}}
\endlgrinde\LGend
}}
\end{minipage}
&&
\begin{minipage}{5.8cm}
{\scriptsize{
\LGinlinefalse \lgrinde
\L{\LB{\K{\#}_\K{class}_\V{ml\_colored\_point}_\V{x}_\V{y}_\V{c}_=}} 
\L{\LB{__\K{object}}}
\L{\LB{_____\K{inherit}}}
\L{\LB{__________\V{callback\_colored\_point}_\V{x}_\V{y}_\V{c}_\K{as}_\V{super}}}
\L{\LB{_____\K{method}_\V{getColor}_\V{()}_\K{=}}}
\L{\LB{__________\N{"ML"}_\^_\V{super}\K{\#}\V{getColor}_\V{()}}}
\L{\LB{__\K{end}\K{;;}}}
\L{\LB{\N{class_ml\_colored\_point_:}}}
\L{\LB{_____________int_-\>_int_-\>_string_-\>_csColoredPoint}}
\L{\LB{\K{\#}} \K{let} \V{ml\_cp}_=}
\L{\LB{____\K{new}_\V{ml\_colored\_point}_\N{8}_\N{9}_\V{"red"}\K{;;}}}
\L{\LB{\N{val_ml\_cp_:_ml\_colored\_point_=_\<obj\>}}}
\L{\LB{\K{\#}_\V{ml\_cp}\K{\#}\V{toString}_\V{()}\K{;;}}}
\L{\LB{\N{- : string = "(8,9):MLred"}}}
\endlgrinde\LGend
}}
\end{minipage}\\ \hline
\end{tabular}
\end{center}

\begin{figure}[hbt]
\begin{center}
\epsfig{file=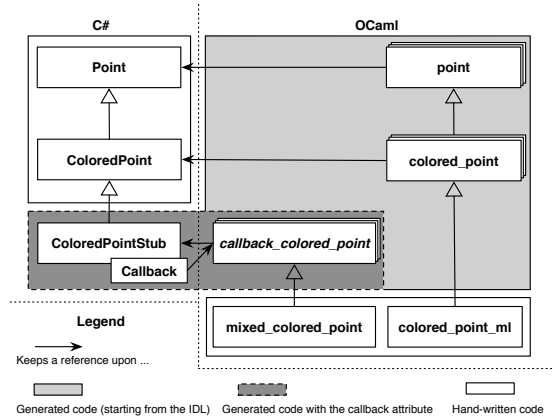,height=0.45\linewidth}
\end{center}
\caption{Relationship between classes}
\label{fig-rel}
\end{figure}

How is this achieved ? 
The two stubs in {\CSHARP} and {\OC} own a reference upon each other.
The {\CSHARP} stub overrides each method as a callback 
to {\OC}, and the {\OC} stub define each method as a non-virtual call
to {\tt ColoredPoint}, the base-class of {\tt ColoredPointStub}. See figure 
\ref{fig-rel} for the complete class diagram.

\forget{
The program {\tt{test\_p.ml}} of figure \ref{fig-exe} shows how to use the {\tt{Point}} and {\tt{ColoredPoint}} classes from {\OC}. 


The left side of figure \ref{fig-exe} corresponds to the basic level of communication whereas 
the right side corresponds to the richer level.
The third call to the {\tt{display}} method illustrates the limitation of the basic level 
since the overriding of {\tt getColor} in {\OC} is ignored by {\CSHARP}.
On the other side, callback mechanism allows communication to be achieved in both ways.

The construction of the list {\tt{l}}, of type {\tt{point list}}, shows the combination of functional and object styles in {\OC} 
involving encapsulated {\CSHARP} objects. 
}

\subsection{Safety considerations}

{\OC} ensures execution safety by static typing,
but what happens when it uses foreign pieces of code ?
We distinguish two kinds of errors:

\begin{itemize}
\item Runtime errors during {\CSHARP} code execution (dynamic cast errors for 
instance), that are not a consequence of inter-operation. 
\item Inconsistency between the {\IDL} and the implementation.
For example, {\CSHARP} components described in the {\IDL} may not be available 
at runtime, or incorrectly described.
\end{itemize}

The first class of errors will fortunately raise runtime exceptions.
They can be considered as ``normal'' runtime errors.
They can be caught by the {\CSHARP} component or, by default, by {\OC} code itself.
As {\OC} exceptions and {\CSHARP} exceptions are both compiled to exceptions
of the underlying runtime, they can easily cross languages boundaries. 

The second class of errors is a consequence of inter-operation itself.
We choose to detect those errors very soon.
{\OC} programs that incorrectly use foreign components 
are detected by static typing at compile time; 
type checking is done with the assumption that {\IDL} types as correct.
This hypothesis can only be checked at runtime with the reflection mechanism.
The code which is generated by {\OJACARED} performs tests at startup time,
immediately acknowledging the programmer of mismatches between components and {\IDL} files.


\forget{
Static type checking ensures that: 

That methods described in the {\IDL} will be called on a ``correct'' instance,
and that their arguments are of the expected type.
I mean, object-arguments passed to an encapsulated {\CSHARP} method by {\OC},
are {\CSHARP}-encapsulated object from expected type, 
and not an pure-{\OC} object, and vice-versa.

We ensure that with typing reference on object of the foreign world,
with an ``abstract'' type (value from those type can't be created outside
the generated code).
}

\subsection{A few words about {\OC} in {\CSHARP}}
In order to call {\OC} methods from {\CSHARP} we reuse the technology behind 
calbacks from {\CSHARP} to {\OC} (see subsection \ref{callback-section}).
Basic encapsulation can similarly be extended with a callback mechanism.

Let us stress the lack of symmetry between {\OC} and {\CSHARP} from an implementation point of view.
Whereas {\CSHARP} objects are directly compiled to CTS objects, {\OC} objects are not: the current implementation
provides its own mechanisms of inheritance and late binding.
Calls from {\OC} to {\CSHARP} directly use reflection mechanisms provided by the {\DOTNET} runtime
but calls from {\CSHARP} to {\OC} have to deal with the peculiarities of {\OC} implementation.
A future release of {\CAMIL} may solve this problem.

As {\OC} does not offer any introspection mechanism for type information, we need
to adapt our dynamic checking.
Even with basic encapsulation, we generate some {\OC} code, that will
statically check {\IDL} against {\OC} implementation, and dynamically check it against
the generated {\CSHARP} at startup time.

\section{Expressiveness and limitations}
The introductory example of section \ref{introducing-section} only involved a simple form of communication.
In this section we try to go a little bit further.
We first give a positive result about the expressiveness of the blending of two different object models.
Then we apply {\OJACARED} technology to a real example that involves complex communication.
It is used as the starting point of a discussion about the actual limitations of {\OJACARED}.

\subsection{Combining the two Objects Models}

{\OJACARED} allows to partially handle both object models. 
We illustrate these new possibilities by showing a case of multiple inheritance in {\OC} of {\CSHARP} classes
and an example of dynamic type checking ({\em downcast}) in {\OC}.


\subsubsection{Multiple inheritance of {\CSHARP} classes}

The following example is taken from \cite{chailloux-manoury-pagano:2000}. 
We define two class hierarchies in {\CSHARP}: graphical objects and geometrical objects.
Each class hierarchy has a class {\tt{Rectangle}}. 
The following {\OC} program defines a class inheriting both {\CSHARP} classes. 
 

\begin{center}
\begin{tabular}{|c|c|}\hline
The file {\tt{rect.idl}} & The {\OC} program \\ \hline
\begin{minipage}{5.3cm}
\vspace{2pt}
{\scriptsize{
\LGinlinefalse \lgrinde
\L{\LB{\K{package}_\V{mypack};}}
\L{\LB{}}
\L{\LB{\K{class}_\V{Point}_\{}}
\L{\LB{_[\K{name}_\V{point}]_\K{\<init\>}_(\K{int},_\K{int});}}
\L{\LB{\}}}
\L{\LB{\K{class}_\V{GraphRectangle}_\{_}}
\L{\LB{_[\K{name}_\V{graph\_rect}]_\K{\<init\>}(\V{Point},_\V{Point});}}
\L{\LB{_\K{string}_\V{toString}();}}
\L{\LB{\}}}
\L{\LB{\K{class}_\V{GeomRectangle}_\{}}
\L{\LB{_[\K{name}_\V{geom\_rect}]_\K{\<init\>}(\V{Point},_\V{Point});}}
\L{\LB{_\K{double}_\V{compute\_area}();}}
\L{\LB{\}}\Tab{3}{}}
\endlgrinde\LGend
}}
\end{minipage}
&
\begin{minipage}{6.3cm}
{\scriptsize{
\LGinlinefalse \lgrinde
\L{\LB{\K{open}_\V{Rect};;}}
\L{\LB{}}
\L{\LB{\K{class}_\V{geom\_graph\_rect}_\V{p1}_\V{p2}_=_}}
\L{\LB{\K{object}}}
\L{\LB{}\Tab{2}{\K{inherit}_\V{geom\_rect}_\V{p1}_\V{p2}_\K{as}_\V{super\_geo}}}
\L{\LB{}\Tab{2}{\K{inherit}_\V{graph\_rect}_\V{p1}_\V{p2}_\K{as}_\V{super\_graph}}}
\L{\LB{\K{end};;}}
\L{\LB{}}
\L{\LB{\K{let}_\V{p1}_=_\K{new}_\V{point}_\N{10}_\N{10};;}}
\L{\LB{\K{let}_\V{p2}_=_\K{new}_\V{point}_\N{20}_\N{20};;}}
\L{\LB{\K{let}_\V{ggr}_=_\K{new}_\V{geom\_graph\_rect}_\V{p1}_\V{p2};;}}
\L{\LB{\V{Printf}.\V{printf}_\3\V{area}=\%\V{g}\2\V{n}\3_(\V{ggr}\V{\#compute\_area}_());;}}
\L{\LB{\V{Printf}.\V{printf}_\3\V{toString}=\%\V{s}\2\V{n}\3_(\V{ggr}\V{\#toString}_());;}}
\endlgrinde\LGend
}}
\end{minipage} \\ \hline
\end{tabular}
\end{center}


\subsubsection{Downcasting {\CSHARP} objects in {\OC}.}
{\OC} does not allow any dynamic typing operations on objects, however inter-operating with {\CSHARP} makes them necessary,
at least for objects coming from a computation on {\CSHARP} side.
The example below builds a list {\tt{l}} of {\tt{csPoint}} objects, even though these actually are colored points.
For each {\CSHARP} class hierarchy described in an {\IDL} file, {\OJACARED} generates a {\OC} class hierarchy, 
which root class is denoted by {\tt{top}}.
{\OJACARED} also generates type coercion functions from {\tt{top}} to the {\OC} type of a {\CSHARP} class. 
These functions raise an exception in case of type inadequacy.

\begin{center}
\begin{tabular}{|ll|} \hline
&
\begin{minipage}{9cm}
\vspace{2pt}
\LGinlinefalse \lgrinde
\L{\LB{\K{let}_\V{l}_=_[(\V{ml\_cp}_:\>_\V{csPoint});_(\V{wml\_cp}_:\>_\V{csPoint})];;}}
\L{\LB{\N{val l : csPoint list = \<obj\>}}}
\L{\LB{\K{let}_\V{lc}_=_\V{List}.\V{map}_(\K{fun}_\V{x}_\-\!\>_\V{csColoredPoint\_of\_top}_(\V{x}_:\>_\V{top}))_\V{l};;}}
\L{\LB{\N{val l : csColoredPoint list = \<obj\>}}}
\endlgrinde\LGend
\vspace{2pt}
\end{minipage} \\ \hline
\end{tabular}
\end{center}





\subsection{Application: a Ray-tracer Program}

We illustrate inter-operability on the following example:
extending an {\OC} program with a graphical interface written in {\CSHARP}.
We use the winning entry of the ICFP'2000 programming contest\cite{raytracer:2000}
which implements a ray-tracer in {\OC}.
Let us state the problem: 
\begin{itemize}
\item The {\OC} class \verb+Render+ defines a {\tt compute} method.
This method expects a string (the name of a file that represents the 3D scene to draw) and a class \verb+Display+ to render pixels on (thanks to calls to a so-called {\tt drawPixel} method).
\item The graphical interface is a class {\tt Display} inheriting from (or holding a reference to an instance of)
 the root widget \verb+System.Windows.Forms.Form+ of {\DOTNET} windowing API, 
with a {\tt drawPixel} method. 
A file dialog helps selecting a 3D scene.
\end{itemize}
Communication is round tripping between the two components.
This can be implemented with {\OJACARED} using cross-language late binding.
Two solutions work:

\begin{figure}
\begin{center}
\epsfig{figure=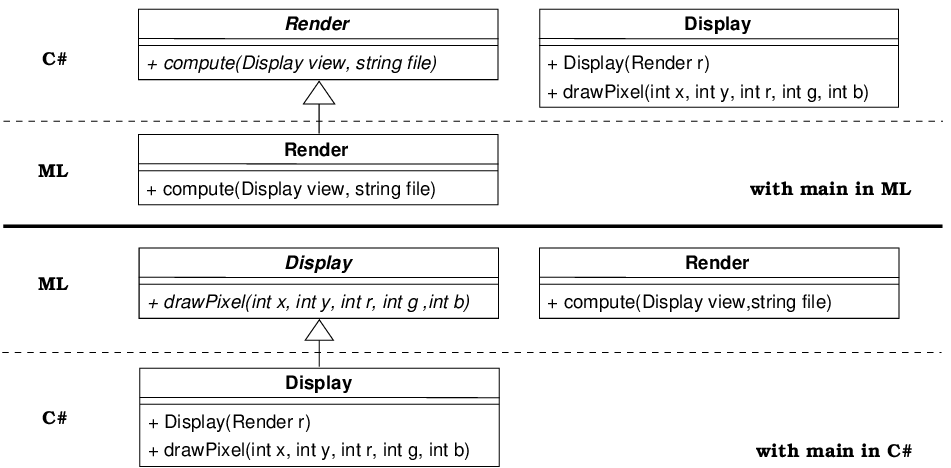,height=0.4\linewidth}
\end{center}
\end{figure}

\begin{enumerate}

\item
The {\CSHARP} interface is parameterized on an abstract class {\tt Render} which defines a {\tt compute} method.
The interface constructor expects an instance of this class. 
When the user chooses a 3D scene file, {\tt compute} is called.
The {\OC} program implements the class {\tt Render}. 
The \verb+Main+ method is on {\OC} side: it passes an instance of {\tt Render} to the constructor of {\tt Display} (starting the graphical interface).
When called, the {\tt compute} method calls {\tt drawPixel} for each computed pixel.

\item
The \verb+Main+ method is on {\CSHARP} side: it builds an instance of {\tt Display} which specializes an abstract class defined in {\OC}.
Here, because multiple inheritance is not allowed in {\CSHARP}, 
{\tt Display} only holds a reference to an object that inherits \verb+System.Windows.Forms.Form+.
When a scene file is selected, it builds a {\tt Render} object and calls {\tt compute} with the filename and \verb+self+.


\end{enumerate}

\begin{center}
\epsfig{figure=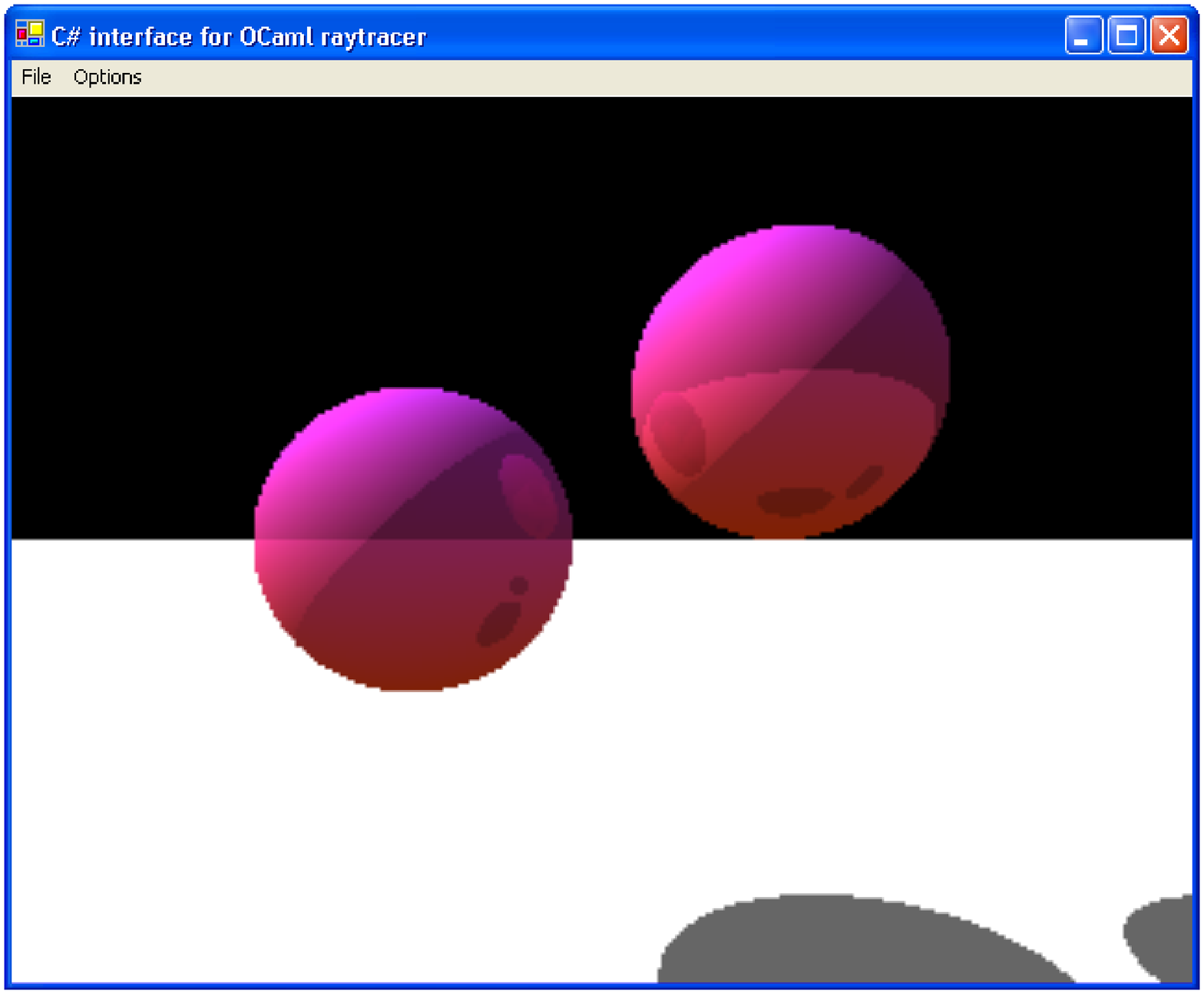,height=0.50\linewidth}
\end{center}

\subsection{How perfect is the blending ?}
\label{blending}
{\OJACARED} allows to use components from one language to another, in both ways.
However we still cannot claim that this can reduce the two worlds into a single one. 

Let us go on with the ray-tracer program.
If we had a single world, we could use {\IDL} files to declare the \verb+drawPixel+ method from the {\CSHARP} class \verb+Display+
and the \verb+compute+ method from the {\OC} class \verb+Render+, making them accessible to both components without using the ``trick''
of redefinition of abstract classes.
Unfortunately, {\IDL} files can only be used to expose classes from one language to the other one.

We cannot simulate one world with two {\IDL} files, 
one for describing the {\CSHARP} {\tt Display} class, 
then one for the {\OC} {\tt Render} class because those classes
are mutually recursive.
The point is:
the \verb+compute+ method expects an instance of \verb+Display+, 
so the latter {\IDL} needs to describe the {\OC} wrapper for the 
{\tt Display} generated from the first {\IDL}.
This leads to typing errors at {\CSHARP} compile time,
because there is no inheritance relationship between the original
{\tt Display} and the twice encapsulated {\tt Display}.

\section{Conclusion and further work}
Our approach differs from {MLj}\cite{MLj},  {SML.NET}\cite{smlnet} and {\FSHARP}\cite{syme01ilx}   projects
which embed {\JAVA} or {\CSHARP} object models inside ML dialects.
We do not modify {\OC} at all, keeping the specificities of its object model.
This leads to a richer model that combines  {\OC}  polymorphisms with {\CSHARP} dynamic typing.

Each community can use {\OJACARED} to import components from the other one. 
However {\OJACARED} needs to be improved.
Some interesting features from the {\DOTNET} runtime (such as methods delegates, genericity, \ldots) should be addressed 
and made available in the {\IDL}.
By making the {\IDL} closer to the CTS, one can also imagine to solve the problem discussed in subsection \ref{blending}.

\forget{
Therefore, with {\OJACARED} we could imagine export some {\OC} objects as
a .NET component, and to use some .NET component into {\OC}.
The {\IDL} could be seen as a simple description of CLR object model,
but should perhaps be extended to be expressive enough.
}

\forget{
The full paper will extend this abstract,  by
giving a complete description of implementation and detailing communication between the two languages,
giving new multi-paradigms programming examples (partial method calls, class instanciation using functors application \ldots), 
comparing benchmarks to discuss about performances and
comparing other approaches as  Lambada \cite{Meijer-Sigbjorn:2001} for Haskell.
} 
\bibliographystyle{splncs}
\bibliography{bibmpool}

\end{document}